\documentclass[showpacs,preprintnumbers,amsmath,amssymb,nofootinbib,superscriptaddress]{revtex4}
\usepackage{graphicx}
\usepackage{epsfig}
\usepackage{bm}
\usepackage{amsfonts}

\newcommand{\eq}{\begin{eqnarray}}
\newcommand{\eqx}{\end{eqnarray}}
\newcommand{\ba}{\begin{equation}}
\newcommand{\ea}{\end{equation}}
\newcommand{\f}[2]{\frac{#1}{#2}}

\newcommand{\dl}{\delta}
\newcommand{\lam}{\lambda}

\newcommand{\ep}{\epsilon}
\newcommand{\bit}{\begin{itemize}}
\newcommand{\eit}{\end{itemize}}
\newcommand{\ii}{\item}

\def\la{\label}
\def\nn{\nonumber \\}
\def\bi{\bibitem}

\def\d{\partial}
\def\th{\theta}

\def\eq#1{{Eq.~(\ref{#1})}}

\reversemarginpar

\begin{document}

\title{Exact $(1\!+\!1)$\,-dimensional flows of a perfect fluid}

\author{Robi~Peschanski}
 \email{robi.peschanski@cea.fr}
 \affiliation{Institut de Physique Th\'eorique  CEA-Saclay
F-91191 Gif-sur-Yvette Cedex, France}

\author{Emmanuel N. Saridakis}
 \email{msaridak@phys.uoa.gr}
 \affiliation{College of Mathematics and Physics,\\ Chongqing University of Posts and
Telecommunications, Chongqing, 400065, P.R. China }

\begin{abstract}
We present a general solution of relativistic
$(1\!+\!1)$-dimensional  hydrodynamics for a perfect fluid flowing
along the longitudinal direction as a function of time, uniformly
in transverse space. The Khalatnikov potential is expressed as a
linear combination of two generating functions with polynomial
coefficients of $2$ variables. The polynomials, whose algebraic
equations are solved, define an infinite-dimensional basis of
solutions. The kinematics of the $(1+1)$-dimensional flow are
 reconstructed from the potential.
\end{abstract}

 \maketitle

\section{Introduction}
\subsection{Historical perspective}
The problem of solving the $(1+1)$-dimensional flow of  a
relativistic perfect fluid has a quite long history in particle
physics. It has been first investigated in the pioneering work
\cite{Lan1} where relativistic hydrodynamics has been introduced
for describing high-energy multiparticle scattering. Together with
the other pioneering work of Ref.\cite{bj} they are considered as
the founding papers of  the modern applications of hydrodynamics
to heavy-ion collisions.

Ref.\cite{Lan1} has been  followed by  studies on the same
guideline \cite{khal,
bi,BelenkijMilekhin,Ros,old,Amai1957,milekhin}.  The Gaussian
rapidity dependence prediction for the ``Landau flow''  found in
\cite{Lan1} consistent with the observed multiplicity
distributions has inspired subsequent works
\cite{Carruthers,(1+1),Chiu}. It has been revived recently
\cite{stein,Wong:2008ex,Mizo} in connection with the experimental
results on heavy-ion collisions at ultra-high energies
\cite{Rhic}.

The other well-known pioneering work analyzing the
$(1+1)$-dimensional  flow of a relativistic perfect fluid is thus
Ref.\cite{bj} (with a precursor \cite{Hwa}). Here, the
boost-invariant solution of the $(1+1)$-dimensional flow, the
``Bjorken flow'', allows for quantitative predictions valid for
the central rapidity region of heavy-ion reactions. It provided a
firm theoretical basis for the prediction of the Quark-Gluon
Plasma produced in subsequent heavy-ion colliders. In fact, it is now
realized that  the flow of
relativistic particles created by the collisions can be well
described by hydrodynamics, at least during some intermediate
stage of the reaction where one observes the creation of a
specific phase of Quantum Chromodynamics, namely the Quark-Gluon
Plasma (QGP) \cite{QGP}.

Recent works on the hydrodynamic behavior of the QGP
\cite{Hydr} uses numerical simulations of hydrodynamics, with the
aim of solving them in a realistic way, including
4-dimensionality of space-time, initial and final
conditions of the hydrodynamic regime, viscosity and other
transport coefficients, realistic equation of state.
However, it is useful to reconsider the initial \cite{Lan1,bj}
problem, namely finding the exact analytic solutions of
hydrodynamic equations in the simplified set-up of a perfect fluid
flow in the longitudinal direction with constant speed of sound.
As we shall see, this problem has not yet been solved.

There are quite a few motivations to follow this path, besides
being the missing piece of a long lasting theoretical physics problem.
On the phenomenological ground, it is known
that in a  first stage  (important for later evolution, as discussed
 already in the seminal papers \cite{Lan1,bj}), the hydrodynamic
flow is mainly (1+1)-dimensional, $i.e.$ can essentially be
described in the kinematic relativistic subspace defined by
proper-time $\tau$ and space-time rapidity $\eta.$

On a more theoretical ground, the recently found Gauge/Gravity
connection \cite{Son,Jani,Minw} between  relativistic
hydrodynamics and gravity in an higher-dimensional space through
the AdS/CFT correspondence motivates  completing the  study of
exact solutions of hydrodynamical equations. For instance in
$(1\!+\!1)$ dimensions, the `Bjorken flow''  of a perfect
fluid in a strongly coupled gauge theory is put in one-to-one
correspondence  \cite{Jani} with the time-dependent 5-dimensional
gravity configuration of a Black Hole  escaping  away in the fifth
dimension. Going beyond the ``Bjorken flow'' is  an important open
question for the application of AdS/CFT correspondence to plasma
physics. Hence, making progress in the exact solution of the
hydrodynamic equations in $(1+1)$ dimensions may be quite useful
in a modern perspective.
\subsection{Position of the problem}
The  state of the art we have to begin with is the following. The
hydrodynamic equations are $a\ priori$ non-linear and as such are
difficult to handle exactly through analytic methods. Only few
particular exact solutions have been found. Apart the noticeable
contributions of the pioneering studies, namely the analytic
asymptotic solution of \cite{Lan1} (the ``Landau flow'' solution),
and the boost-invariant solution of \cite{bj} (the ``Bjorken
flow'' solution), there were only few interesting
exact solutions given in the literature  for specific values of
the dynamical parameters (see $e.g.$ \cite{Cs1,new,g2,Nagy},
\cite{Pra,g1,Wong,Suzuki:2009re}). To our knowledge, a general
solution for the relativistic $(1\!+\!1)$-dimensional flow of a
perfect fluid is still lacking.

Recently, two developments on exact solutions of the
$(1\!+\!1)$-dimensional flow appeared, which are the building  blocks
of the present work. On the one hand, aone-parameter family of
solutions, interpolating between the ``Bjorken flow'' and the
``Landau flow'' was derived \cite{Bipe}. They were named {\it
harmonic flows} since they are obtained assuming that the physical
rapidity $y$ is an harmonic function of the light-cone kinematic
variables, condition which is  valid both for the
``Bjorken flow'' and the ``Landau flow''. On the other hand, it
was possible using the formalism of the Khalatnikov potential
\cite{khal} to derive exact solutions of the $(1\!+\!1)$-dimensional
entropy flow as a function of rapidity \cite{Beuf}. The
Khalatnikov potential method makes use of a {\it hodograph}
transformation, allowing for a substitution of the kinematic
light-cone variables by the hydrodynamic ones, namely temperature
and rapidity, in order to transform the initially nonlinear
mathematical problem, posed by the hydrodynamic equations, into a
linear one.

In the present  paper we show how, by combining both approaches,
$i.e.$ the ``harmonic flow'' and the Khalatnikov potential
approach, one generates an infinite-dimensional linear basis of
exact solutions, making a  sizable step towards the general
solution of the relativistic $(1\!+\!1)$-dimensional flow of a
perfect fluid.

Our plan is the  following: In section II,  we provide a reminder
on the Khalatnikov potential method \cite{khal} and recall those
results obtained in Refs. \cite{Bipe,Beuf} for the {\it harmonic flow} solution and its  entropy flow which we will use here. In  section
III, we introduce the notion of $regular$ (${\rm resp.}\
irregular$) solutions obtained by integration ($resp.$ derivation)
from the ``harmonic flow'' and give first generic examples of
solutions. Focusing in section IV on $regular$ solutions, we
derive the more general set of solutions by solving appropriate
polynomial equations in two variables. Section V is devoted to a
discussion of the general solution. A final section VI provides a
summary of our results and an outlook on the prospects for a complete solution
of the exact $(1\!+\!1)$\,-dimensional flows of a perfect fluid.

\section{Khalatnikov equation and Harmonic solutions}
\label{Khalequation}

\subsection{Hydrodynamic equations}

We consider a perfect fluid whose  energy-momentum tensor is
\begin{equation}
 T^{\mu\nu}= (\epsilon+p)u^{\mu}u^{\nu} - p \eta^{\mu\nu}  \label{Tmn}
\end{equation}
 where $\epsilon$ is the energy density, $p$ is the
pressure and $u^{\mu}$ ($\mu =\{0,1,2,3\}$) is the  4-velocity in the
Minkowski metric $\eta^{\mu\nu}$. It obeys the  equation
\begin{equation}
\d_\mu T^{\mu\nu}=0\ .
 \la{hydro}\end{equation}
We write the standard thermodynamical identities  (where we have
assumed for simplicity vanishing chemical potential):
\begin{equation}
p+\epsilon = Ts\;;\;\; d\epsilon = T ds\;;\;\; dp = s dT
\la{therm}\ ,
\end{equation}
where $p,\epsilon,s$ are respectively, the pressure, energy and entropy density. The system of hydrodynamic equations closes  by relating energy
density and pressure through the equation of state, which, in the present study will be considered with constant speed of sound, namely
\begin{equation}
\frac{dp}{d\ep}=\frac{s dT}{T ds}=c_s^2\equiv cnst.
 \label{egpvar}
\end{equation}
We consider now the (1+1) approximation of the hydrodynamic flow,
restricting it only to  the longitudinal direction. Within such an
approximation, the effect of the transverse dimensions is only
reflected through the equation of state \eqref{egpvar}. Note that
we do not $a \ priori$ assume the traceless condition $T^{\mu\mu}=0,$
and thus the fluid is considered as ``perfect'' (null viscosity)
but not necessarily ``conformal'' (null trace).

Let us introduce the light-cone coordinates
\begin{equation}
 z^\pm= z^0\pm z^1 \equiv t\pm z= \tau e^{\pm \eta}\;
\Rightarrow \; \left(\f {\d}{\d z^0}\pm \f {\d}{\d z^1}\right)=2
\f {\d}{\d z^\pm} (\equiv2 \d_\pm)
 \label{dz},
 \end{equation} where
$\tau=\sqrt{z^+z^-}$ is the proper time and
$\eta=\frac{1}{2}\ln({z^+}/{z^-})$ is the {\it space-time
rapidity} of the fluid. We also introduce for further use the
hydrodynamical variables,  namely $y,$  the usual {\it
energy-momentum rapidity} variable and $\theta,$ the logarithm of
the inverse temperature, namely (recalling $u^+u^-=1$)
 \ba
 y= \log u^+ = - \log u^- \ ;\quad\quad \theta= \log ({T_0}/{T})\ ,
\label{uplus}
\ea
 where $u^\pm =\log (u^0\pm u^1)$ are the light-cone
components of the fluid velocity and $T_0$ some given fixed
temperature,  $e.g.$ the initial one for a cooling plasma
(explaining why one choses the ratio of temperatures
\eqref{uplus}, leading to $\theta \ge 0$).

The hydrodynamic equations (\ref{hydro}) take the form
\begin{eqnarray}
 \left(\partial_++
\partial_-\right)T^{00}+\left(\partial_+- \partial_-\right)T^{01}&=&0
\nonumber \\
\left(\partial_++ \partial_-\right)T^{01}+\left(\partial_+-
\partial_-\right)T^{11}&=&0\ .
\label{eqbasic}
\end{eqnarray}
Note that inserting the formulation of the energy-momentum tensor
\eqref{Tmn} into the system \eqref{eqbasic} using  expressions
\eqref{uplus} leads to an highly non linear system of equations in
terms of the kinematic phase-space variables \eqref{dz}. This
explains why there happened to be so much difficulty to
find exact solutions of the flow characteristics. This is our aim
to find a general solution to this problem by a  change of
perspective.

\subsection{The Khalatnikov Equation}
\label{Khalderivation}

It is known \cite{khal,bi} that one can
replace to the non-linear problem of (1+1) hydrodynamic
evolution with a linear equation for a suitably defined potential.
In this section we briefly recall the results of Refs.\cite{khal,bi}  (recasting the
calculations
in the light-cone variables, as was done in \cite{Bipe}).

Using the thermodynamic relations \eqref{therm}, one can recombine
the two equations \eqref{eqbasic} into the following ones, each of
them having a physical interpretation, namely
\bit \ii {\it The
flow derives from a kinematic potential} \eit
One combination of
Eqs.\eqref{eqbasic} gives
\begin{equation}
\partial_+\left(e^{-\th+ y}\right)=\partial_-\left(e^{-\th- y}\right)\equiv \partial_+\partial_-\Phi(z^+,z^-)
.\label{rel1}
\end{equation}
Eq.\eqref{rel1} proves the existence of a potential $\Phi(z^+,z^-)$ such that:
\begin{equation}
\partial_\mp\Phi(z^+,z^-)\equiv u^\pm T=T_0\ e^{-\th\pm y}\ .\label{partialphi}
\end{equation}
\bit
\ii {\it Conservation of entropy}
\eit
Another independent  combination of equations \eqref{eqbasic} corresponds to the conservation of entropy, namely
\begin{equation}
\partial_+\left(u^+s\right)+\partial_-\left(u^-s\right)=0
\label{varentropeq}.
\end{equation}

Combining Eqs.(\ref{partialphi}) and (\ref{varentropeq}), one introduces the {\it Khalatnikov potential}
\begin{equation}
\chi(\theta,y)\equiv\Phi(z^+,z^-)-z^-u^+T-z^+u^-T\ ,
\label{chi0}
\end{equation}
where $z^{\pm}$ are now considered as functions of $(\theta,y)$.
This  is called the {\it hodograph} transformation expressing the
hydrodynamic equations as a function of    the dynamical variables
($\theta,y$) $via$ the Legendre transformation \eqref{chi0}. The
kinematic variables are recovered from the Khalatnikov potential
by the equations
\begin{equation}
z^\pm(\theta, y)=\frac{1}{2 T_0}\ e^{\theta \pm y}\ \left(\partial_\theta\chi
\pm \partial_y\chi\right)\ . \label{zpzm}
\end{equation}

The Khalatnikov potential has the remarkable property \cite{khal,bi,Beuf} to verify a
$linear$ partial differential equation which takes the form\footnote{Note that in this relation there is a sign difference comparing to
that of \cite{Bipe,Beuf}, due to the sign-difference in the
$\theta$-definition that we use in this work.}
\begin{equation}
c^2_s\,\partial_\theta^2
\chi(\theta,y)-\left[1-c_s^2\right]\partial_\theta
\chi(\theta,y)-\partial^2_y \chi(\theta,y)=0\ .
 \label{Khalatnikov}
\end{equation}
In \eqref{Khalatnikov}, $c_s$ (denoted also $1/\sqrt g$ for
further convenience) is the speed of sound in the fluid, which
will be considered as a constant in the present study.

The Khalatnikov equation has been originally  derived \cite{khal}
for the potential \eqref{chi0}. But its range of applicability
appears to be much wider. Indeed, the transformation of a
nonlinear problem in terms of the kinematic variables into a
linear one in terms of hydrodynamic variables has tremendous
advantages, as we shall see further.  It allows
to obtain new solutions by arbitrary linear combinations of known
ones. Furthermore, primitive integrals and derivatives
of solutions are also solutions.

In order to illustrate the powerfulness  of this method, let us
give two already known \cite{Bipe} examples. If $\chi(\theta,y)$
is solution of \eqref{Khalatnikov}, then the potential $\Phi,$
defined through \eqref{partialphi}, but now expressed in terms of
the hydrodynamic variables through \eqref{zpzm}, reads
\begin{equation}
\Phi(\theta,y)\equiv\Phi\{z_+(\theta,y),z_-(\theta,y)\}=
\chi(\theta,y)+\partial_\theta \chi(\theta,y)\equiv
e^{-\th}\partial_\theta \{ e^{\th}\
\chi(\theta,y)\}\label{PhiChi},
\end{equation}
and thus it verifies also the Khalatnikov equation
\eqref{Khalatnikov}. As a direct consequence, the physical entropy
flow as a function of rapidity also verifies \eqref{Khalatnikov}.
Indeed, one has \cite{Beuf} (see also \cite{milekhin})
\begin{equation}
\frac{dS}{dy}(\theta,y) = \frac{s_0}{2 gT_0}\,
e^{-(g-1)\th}\ \partial_\theta \Phi_{} (\theta,y)\
,\label{Entr}
\end{equation}
where  we use the thermodynamic relation $s=s_0 e^{-g\th}$ for the
overall, temperature-dependent, entropy density of a perfect
fluid, recalling that by definition $c_s^2\equiv 1/g$.

Before going further, it proves useful to use new variables, which
allow to put the  Khalatnikov equation \eqref{Khalatnikov} in a
simple and symmetric form.  Introducing
\begin{eqnarray}
 a\equiv\f 12\ {\sqrt{(g\!-\!1)(\theta + c_s{y})}},\ \ \ \
b\equiv\f 12\ {\sqrt{(g\!-\!1)(\theta - c_s{y})}}\
\label{ab}
\end{eqnarray}
and redefining $\chi(a,b)$ as a function of the reduced variables
of \eqref{ab}, and introducing
\begin{equation}
Z(a,b)= e^{-(a^2+b^2)}\chi(a,b),
\end{equation}
then the Khalatnikov equation (\ref{Khalatnikov}) takes one or the
other simple forms
\begin{eqnarray}
\partial_{a^2}\partial_{b^2}\
 \chi(a,b) &=&\{\partial_{a^2}+\partial_{b^2}\}\
 \chi(a,b)
\la{KhalK}\\
\partial_{a^2}\partial_{b^2}\
Z(a,b)&=&Z(a,b) \ .
\label{KhalZ}
\end{eqnarray}

\subsection{Harmonic Flow}
As noticed in \cite{Bipe}, a specific combination of the hydrodynamic
set of equations (\ref{therm},\ref{eqbasic}) allows one to
eliminate the temperature and write a consistency condition on the
rapidity. It reads
 \ba
4\ \d_+\d_- y= \frac{g\!-\!1}{g\!+\!1}\
\left\{\d_-\d_-[e^{-2y}]-\d_+\d_+[e^{+2y}]\right\}\ .
\la{cs}
\ea
Eq. \eqref{cs} explicitly exhibits the highly nonlinear
character of the hydrodynamic equations \eqref{eqbasic} written in terms of
kinematic differentials.

Despite this nonlinear feature, an analytic one-parameter  family
of solutions  interpolating between the Landau and Bjorken flows
has been obtained \cite{Bipe} imposing the harmonic condition
(with $\{+,-\}$ signature) \ba \d_+\d_-\ y\equiv
\{(\d_{t})^2-(\d_z)^2\}\ y=0\ . \la{Harmo} \ea Harmonicity and the parity
symmetry by $z_\pm$ interchange can be realized by writing \ba
y(z_+,z_-)= \f 12 \left[l^2(z_+) -l^2(z_-)\right]\ , \la{Harm} \ea
with \ba
 z_\pm = \int^{l^{\pm}}{dl}\ \ e^{l^2}
\la{Solh}
\ea
as an implicit equation defining $l(z_\pm)$  as
a function of the kinematics (up to constants).

The relation between thermodynamic and kinematic variables can then be
explicitly written
\begin{eqnarray}
\theta &=&
\frac{g+1}{4g}\left({l_+}^2+{l_-}^2\right)-\frac{g\!-\!1}{2g}\
l_+
l_-\nonumber\\
y &=& \frac{1}{2} \left({l_+}^2-{l_-}^2\right)\ ,\label{thetaHarm}
\end{eqnarray}
and reversely by simple algebraic manipulation
\begin{eqnarray}
l_+ + l_- &=& \sqrt{2g}\
\left(\theta+\sqrt{\theta^2-{y^2}/{g}}\right)^{1/2}=\sqrt{\f{{4g}}{g\!-\!1}} \ (a+b)\nonumber\\
l_+ - l_- &=& \ \sqrt{2}\
\left(\theta-\sqrt{\theta^2-{y^2}/{g}}\right)^{1/2}= \sqrt{\f{ 4}{g\!-\!1}}\ \ (a-b)
\label{lpluslminus}
,
\end{eqnarray}
where we introduce the reduced variables \eqref{ab}. Note that each of the  variables $a,b$ is a function of both
$z_\pm,$ contrary to the ``harmonic variables'' $l_+(z_+),l_-(z_-).$

Using  the property \eqref{partialphi} of the potential $\Phi$ one writes
\begin{equation}
\frac{\partial \Phi}{\partial l_\pm} = \frac{d z^\pm}{d l_\pm} \ \partial_\pm
\Phi = \frac{d z^\pm}{d l_\pm}\  T_0 \ e^{-\theta \mp y}\ .
\label{1}\end{equation}
Now, inserting \eqref{Harm} and  \eqref{thetaHarm}, one
obtains
\begin{eqnarray}
\frac{\partial \Phi}{\partial l_\pm} &=& \ e^{l_\pm^2} e^{-\theta
\mp y}=  \ e^{\frac{g\!-\!1}{4g} (l_+ + l_-)^2}\ .\label{2}
\end{eqnarray}
The expression \eqref{2} is symmetric in $l_\pm$  and thus, by mere
integration and using \eqref{lpluslminus}, one gets for the kinematic potential
\begin{equation}
\Phi(a,b)\propto\ \int^{a+b}_c  dt\ e^{t^2}\ ,\label{PhiHarm0}
\end{equation}
where the initial integration value $c$ is matter of convention.

Hence the harmonic flow derives from a simple potential
which corresponds to is the  ``Imaginary Error Function'' defined as
\begin{equation}
{\rm{erfi}}[z]=\frac{2}{i\sqrt{\pi}}\int_0^{iz}e^{-t^2}dt=\frac{2}{\sqrt{\pi}}\int_0^{z}e^{t^2}dt\ .
\end{equation}
Using   relation \eqref{Entr} and the definitions \eqref{ab}
one easily obtains the  entropy distribution corresponding to the
harmonic flow
\begin{equation}
\frac{dS}{dy} = \frac{s_0}{2 gT_0}\,e^{-(g-1)\th}\,
\partial_\theta \Phi_{} (\theta,y)\propto \f {a+b}{ab}\ e^{-(a-b)^2}\ .\label{dS_harmonic}
\end{equation}
As noticed in \cite{Bipe}, this entropy distribution, considered
for freeze-out  at a fixed proper-time, leads to a density which
interpolates between the Landau Gaussian solution and the Bjorken
boost-invariant one. However the distribution contains a
singularity at $a,b=0$, $i.e.$ when $c_s y\to\pm\theta,$ which
causes a phenomenological problem. We will see that besides
$irregular$ solutions for the entropy flow generalizing the one
obtained for the harmonic flow \eqref{dS_harmonic}, a full set of
$regular$ solutions will be found in the set of $1\!+\!1$ flow
solutions, thus avoiding the phenomenological problems of the
harmonic flow.

\section{Regular and irregular solutions}
\subsection{Derivatives of the harmonic flow}

Thanks to the linear form of Khalatnikov equation \eqref{Khalatnikov}, it is obvious that any
derivative of $\chi$ with respect to $y$ or/and $\theta$ will also provide
 a solution. In terms of the reduced variables \eqref{ab}, the derivatives of $\chi$ with respect to
$a^2$ and $b^2$ (which are linear in $y$ and $\theta$) will be
also  solutions of the Khalatnikov equation. Hence, this is also
valid for the potential $\Phi(a,b)$, thanks to the general linear
relation \eqref {PhiChi}. Note that symmetric derivatives in $a^2$
and $b^2$ will correspond to symmetric solutions in rapidity which
we keep studying in the present paper.

Let $\Phi^{(h)}_{n}$ denote the $n$-th derivative of the harmonic
potential \eqref{PhiHarm0} with respect to $a^2$ and $b^2$. After
some  algebra, one realizes that the general solution for
$n\geq1$ is the product of an exponential with a rational fraction
of symmetric polynomials in $a$ and $b$
\begin{equation}
\Phi^{(h)}_{n}(a,b)\equiv(\d_{a^2})^n(\d_{b^2})^n\Phi^{(h)}(a,b)=e^{(a+b)^2}\frac{Q_{n}(a,b)}{R_{n}(a,b)}\
,
\end{equation}
where the denominator takes the form
\begin{eqnarray}
 R_{n}(a,b)&=&a^{n}b^{n}\ ,
\label{Deriv}
\end{eqnarray}
and $Q_{n}(a,b)$ is a polynomial which can be straightforwardly
determined through the iteration of derivatives. In fact, unless
very particular cases (we did not find a counter-example)  the
general derivatives of the harmonic solution  possess multiple
poles at $a=0$ and $b=0$ , $i.e$ when $y=\pm\sqrt{g}\,\theta$ which
appear as singularities in the entropy distribution \eqref{Entr}.
We call them {\it irregular} solutions since they lead to
singularities in a physical distribution, the
first example being the  single poles of the harmonic  flow
itself, see \eqref{dS_harmonic}.

Finally, note that all derivative solutions depend only on one
non-meromorphic function $ e^{(a+b)^2},$ which we call the {\it
seed function} since all derivatives come from and factor out this
function.

\subsection{Integrals of the harmonic flow}

Integrals of the harmonic flow potential \eqref{PhiHarm0} verify
the Khalatnikov equation \eqref{Khalatnikov}, provided one takes
care of the boundary conditions (this amounts to keep $c=0$ in
\eqref{PhiHarm0}).  Let $\Phi^{(h)n}$ denote the $n$-th integral
with respect to $a^2$ and $b^2$. Thanks to the mathematical
property of the error function
\ba
\int{\rm{erfi}}[z]dz=z{\rm{erfi}}[z]-e^{z^2}/\sqrt{\pi}\ .
\la{seed}
 \ea
One realizes the  interesting novelties
of the {\it integral} solutions with respect to the {\it
derivative} solutions, namely :
 \bit
  \ii
  The solution contains two transcendental
{\it seed functions}, namely ${\rm{erfi}}[a+b]$ and $e^{(a+b)^2}$, instead of only the last one.
 \ii By iteration of formula
\eqref{seed} and appropriate integrations by parts, the solutions
are always combinations of the two seed functions with
polynomials.
 \ii The solutions are regular\footnote{However higher-order derivatives may be singular.} at $a=0$ and $b=0$. \eit

Indeed, using the reduced variables \eqref{ab}
 for a
more economic notation, one obtains the general form:
\begin{equation}
\Phi^{(h)n}(a,b)=\left[\Pi^{(0)}_{n}(a,b)\Phi_0(a,b)
 +\Pi^{(1)}_{n}(a,b)\Phi_1(a,b)\right],
\end{equation}
where
\begin{equation}
\Phi_0(a,b)=\, e^{
 (a+b)^2}\ ;\quad
\Phi_1(a,b)=
\frac {\sqrt{\pi}}2\ {\rm{erfi}}\left(a+b\right)\equiv \int_0^{a+b}e^{t^2}dt\ ,
 \label{P1}
\end{equation}
and $\Pi^{(0)}_{n}(a,b),\ \Pi^{(1)}_{n}(a,b)$ are symmetric
polynomials in $(a,b)$.  Hence $\Phi^{(h)n}$ does not possess
poles.

As we shall see in the next section, there is a general derivation
of these regular solutions which  will give practical access to
the infinite set of polynomials $\Pi^{(0)}_{n}(a,b),\
\Pi^{(1)}_{n}(a,b).$ However as a first example we provide the
first polynomials, explicitly after 1 and 2 integrations
symmetrically in $a\ {\rm and}\ b$:
\begin{eqnarray}
&&\Pi^{(0)}_{1}=2a^3-2ba^2-a+(a\leftrightarrow b)\nonumber\\
&&\Pi^{(1)}_{1}=-4a^4+ 4a^2(1+b^2)+\f 12+(a\leftrightarrow b),
\la{int1}\end{eqnarray}
\begin{eqnarray}
&&\Pi^{(0)}_{2}=-2[a^7-ba^6-a^5(22+3b^2)+a^4(18b+3b^2)+a^3(36+4b^2)+12ba^2+72a]+(a\leftrightarrow b)\nonumber\\
&&\Pi^{(1)}_{2}=2\{a^8-a^6(24+4b^2)+a^4(72+24b^2+3b^4)+a^2(96+24b^2)+72+(a\leftrightarrow
b)\} \la{int2}.\end{eqnarray} Note that, if $\Pi^{(i)}_{n}$ is
solution also is $\left(\lambda\ \Pi^{(i)}_{n}+\mu\right)$
 for any $\lambda,\mu$ being constants, that is independent of $i,n.$

\section{General regular solution}
Let us now present a systematic way  to find the general solutions
for the integral case. Using the form \eqref{KhalZ} of the
Khalatnikov equation, let us propose a general Ansatz for the
solution having the form
\begin{equation}
Z(a,b)=\left[P^{(0)}(a,b)\ Z_0(a,b)
 +P^{(1)}(a,b)\ Z_1(a,b)\right],
 \label{Znm}
\end{equation}
where
\begin{eqnarray}
&&Z_0(a,b)= e^{2ab}
 \label{Z0}\\
 &&Z_1(a,b)=e^{-a^2-b^2}\int_0^{a+b}e^{t^2}dt,
 \label{Z1}
\end{eqnarray}
and $P^{(0)}$, $P^{(1)}$ are functions determined in such a way
that  \eqref{KhalZ} be satisfied.

 Using the relations:
\begin{eqnarray}
&&\partial_{a^2}Z_0= Z_0\frac{b}{a}\quad\quad\quad\quad\quad\partial_{b^2}Z_0= Z_0\frac{a}{b}
 \nonumber\\
&&\partial_{a^2}Z_1= -Z_1+\frac{Z_0}{2a}\quad\quad\ \partial_{b^2}Z_1=-Z_1+\frac{Z_0}{2b}
 \nonumber\\
&&\!\!\!\!\!\!\!\!\partial_{a^2}\partial_{b^2}Z_1= Z_1\quad\quad\quad\quad\ \partial_{a^2}\partial_{b^2}Z_0= Z_0+\frac{Z_0}{2ab}\ ,
\label{relations}
\end{eqnarray}
we find that the Khalatnikov equation (\ref{KhalZ}) breaks into
two coupled equations, namely:
\begin{eqnarray}
&&\{\partial_{a^2}+\partial_{b^2}-\partial_{a^2}\partial_{b^2}\}P^{(1)}=0\nonumber\\
&&\{a\partial_{a}+b\partial_{b}+1
+\frac{1}{2}\partial_{a}\partial_{b2}\}P^{(0)}=-\f12\{\partial_{a}+\partial_{b}\}P^{(1)}\ .
\label{Eqns}
\end{eqnarray}
In the following we will consider only ``symmetric'' solutions in
the interchange  $a\leftrightarrow b,$ but the method can be of
more general validity.

Let us now for convenience use the
variables:
\begin{eqnarray}
u=a^2+b^2\equiv \f {g\!-\!1}2\ \theta \quad\quad
v=a^2-b^2\equiv \f {g\!-\!1}{2\sqrt g}\ y
\label{vari},
\end{eqnarray}
thus $\partial_{a^2}=\partial_u+\partial_v$ and
$\partial_{b^2}=\partial_u-\partial_v$. In terms of these
variables, equation (\ref{Eqns})  becomes:
\begin{eqnarray}
\label{eqn1}
\left[\partial^2_u-\partial^2_v-2\partial_u\right]P^{(1)}(u,v)=0.
\end{eqnarray}
As a trial, since we expect $P^{(1)}$ to be a  polynomial in two variables
 $(u,v)$,  which  in terms of $(a,b)$ is symmetric (thus
containing only even powers of $v$), we consider the expansion:
\begin{eqnarray}
\label{exp1}
P^{(1)}(u,v)=\sum_{k=0}^{K}\,v^{2k}P_{k}^{(1)}(u),
\end{eqnarray}
where, for an arbitrarily chosen maximal value of the index
$k_{max}\equiv K$, $P_{K}^{(1)}(u)$ are functions
 of $u$. We shall verify later on that they are
 indeed well-defined {\it polynomials}.

At this stage we  introduce the Laplace transform of the functions
$P_{k}^{(1)}(u)$ as:
\begin{eqnarray}
\label{Laplacetransform1}
&&\widetilde{P}^{(1)}_k(\lambda)=\int_{0}^\infty du\,
e^{-\lambda u}\,P_{k}^{(1)}(u)\nonumber\\
&&P^{(1)}_k(u)=\int^{\lambda_0+i\infty}_{\lambda_0-i\infty}\frac{d\lambda}{2\pi
i}\, e^{\lambda u}\
\widetilde{P}^{(1)}_k(\lambda)\ ,\label{Laplacetransform2}
\end{eqnarray}
where $\lambda_0$ is some positive real constant at the right of
all singularities of the integrand.  Therefore, insertion of
(\ref{Laplacetransform2}) into (\ref{exp1}) and then into
differential equation (\ref{eqn1}) gives:
\begin{equation}
\int^{\lambda_0+i\infty}_{\lambda_0-i\infty}\frac{d\lambda}{2\pi
i}\ e^{\lambda
u}\sum_{k=0}^{K}\left[\lambda(\lambda-2)v^{2k}-2k(2k-1)v^{2(k-1)}
\right]\widetilde{P}^{(1)}_{k}(\lambda)=0.
\end{equation}
Starting with the highest power $v^{2K}$ , all coefficients of the
lower powers of $v^2$ must be zero. As we observe, the highest
power contains only one term, namely the coefficient of $v^{2K},$
which amounts to impose \ba \label{coefK}
\int^{\lambda_0+i\infty}_{\lambda_0-i\infty}\frac{d\lambda}{2\pi
i}\ e^{\lambda
u}\lambda(\lambda-2)\widetilde{P}^{(1)}_{K}(\lambda)\equiv 0\ .
\ea In order to avoid any singularity in the $\lambda$ complex
plane, one gets two (non-trivial) possibilities, namely
\begin{eqnarray}
\label{Laplace1}
\widetilde{P}^{(1)}_K(\lambda)\propto \f 1{\ \ \lam\ \ } \quad &&\Rightarrow\quad {P}^{(1)}_K(u)= const.\\
\widetilde{P}^{(1)}_K(\lambda)\propto \f 1{\lam\!-\!2}\quad &&\Rightarrow\quad {P}^{(1)}_K(u)=const.\ \times\ e^{2u}
\label{Laplace2}.
\end{eqnarray}
This leads {\it a priori} to {\it two  families} of
solutions, depending on the chosen highest degree $K$. As we shall
see further on, only the first family
solution of \eqref{Laplace1} will survive the
system of equations \eqref{Eqns}. The second family  \eqref{Laplace2} will meet an obstruction when trying to solve the second equation of \eqref{Eqns}. Hence only polynomial solutions for ${P}^{(0,1)}$ are allowed.

For all smaller powers of $v^2$ we always have two terms, and the
condition for the coefficients to be identically zero read
iteratively as
\begin{eqnarray}
&&\widetilde{P}^{(1)}_{K\!-\!1}(\lambda)=\ \frac{\ \ 2K(2K\!-\!1)\ \ }{\lambda(\lambda-2)}\ \ \widetilde{P}^{(1)}_{K}(\lambda)\
=\ \frac{\Gamma(2K\!+\!1)}{\Gamma(2K\!-\!1)\lambda(\lambda\!-\!2)}\ \widetilde{P}^{(1)}_{K}(\lambda)
\nonumber\\
&&\widetilde{P}^{(1)}_{K\!-\!2}(\lambda)=\frac{(2K\!-\!2)(2K\!-\!3)}{\lambda(\lambda\!-\!2)}\widetilde{P}^{(1)}_{K-1}(\lambda)=
\frac{\Gamma(2K\!+\!1)}{\Gamma(2K\!-\!3)[\lambda(\lambda\!-\!2)]^2}\ \widetilde{P}^{(1)}_{K}(\lambda)
\nonumber\\
&&\cdots
\end{eqnarray}
which straightforwardly leads (up to a common constant) to \ba
\widetilde{P}^{(1)}_{k}(\lambda)=
\frac{\Gamma(2K+1)}{\Gamma(2k+1)}\ \times\
\f{\!\!\!\!\widetilde{P}^{(1)}_{K}(\lambda)}{[\lambda(\lambda\!-\!2)]^{K\!-\!k}}\
, \la{Iter} \ea with $1/ {\lam}$  (for the family of Eq. \eqref{Laplace1}) or
$1/ {(\lam\!-\!2)}$  (for the family of Eq. \eqref{Laplace2}). Finally, the
inverse Laplace transform (\ref{Laplacetransform2}) gives
\begin{eqnarray}
\label{Pkappa}
P^{(1)}_{k}(u)=\int_{\circlearrowleft}\frac{d\lambda}{2\pi
i}\ e^{\lambda u}\,
\frac{\Gamma(2K+1)}{\Gamma(2k\!+\!1)[\lambda(\lambda\!-\!2)]^{K-k}}\ \widetilde{P}^{(1)}_{K}(\lambda)\ ,
\end{eqnarray}
where the complex integration contour ${\circlearrowleft}$
encircles\footnote{Initially the straight imaginary line contour
of \eqref{coefK} can be deformed and leads to encircle the two
multipole singularities at $\lam=0$ and $\lam=2.$ However, as we
shall see, at each step $k$ of the iteration \eqref{Pkappa}, the
choice $\lam=0$ will be selected by the second equation
\eqref{Eqns}.} either $\lam=0$ or $\lam=2.$  From \eqref{Pkappa},
it is clear enough that first family of solutions are polynomials,
the second family being made of polynomials factors of $e^{2u}.$
\subsubsection{First family (allowed)}

Let us first examine the first family generated at all steps by the
multipole at $\lam=0.$ The method will be to obtain the explicit
solution \eqref{Pkappa} of   the first of  equations \eqref{Eqns}
and then plough the solutions as an input in the second member of
the second equation \eqref{Eqns}.

From \eqref{Pkappa} with $\widetilde{P}^{(1)}_{K}(\lambda)\equiv 1/\lam$, $i.e. \ {P}^{(1)}_{K}(u)\equiv 1,$ one obtains
\begin{eqnarray}
\label{Plam1}
P^{(1)}_{k}(u)=\int_{\circlearrowleft}\frac{d\lambda}{2\pi
i}\ e^{\lambda u}\,
\frac{\Gamma(2K+1)}{\Gamma(2k\!+\!1)[\lambda(\lambda\!-\!2)]^{K-k}}\ \times \f 1\lam\ \equiv \frac{\Gamma(2K\!+\!1)}{\Gamma(2k\!+\!1)\Gamma(K\!-\!k\!+\!1)}\ \left[\partial^{K-k}_\lam\left\{\f {e^{\lam u}}{(\lam\!-\!2)^{K-k}}\right\}\right]_{
\lam=0}\ .
\end{eqnarray}
In order to now introduce the second equation \eqref{Eqns}, it is
convenient to expand the polynomials $P^{(0,1)}$ in terms of their homogeneity components of degree $d$ in the variables $(a,b)$, namely
\begin{eqnarray}
&&{P}^{(1)}(a,b)=\sum_{p=0}^{2K} {P}_d^{(1)}(a,b)\quad\quad\ ;\quad d=2p
\nonumber\\
&&{P}^{(0)}(a,b)=\sum_{p=0}^{2K-1} {P}_d^{(0)}(a,b)\quad\ ;\quad d=2p+1\ ,
\la{Homog}
\end{eqnarray}
where the maximal degrees of homogeneity are dictated by the
expansion \eqref{exp1}  and the expression \eqref{Plam1} for
${P}^{(1)}(a,b)$ and then by the second member of the second
equation \eqref{Eqns} for ${P}^{(0)}.$ Indeed, from \eqref{Plam1}
it is straightforward to realize that the maximal degree at level
$k$ for ${P}^{(1)}$ is $d=4k+2(K-k)=2(k+K).$ Note also that
${P}^{(1)}(a,b)$ has only even degrees while ${P}^{(0)}(a,b)$ only
odd ones.

Inserting the expansions \eqref{Homog} into the inhomogeneous
second equation of \eqref{Eqns},  one finds the following nested
recurrence
\begin{eqnarray}
(4K)\ {P}_{4K-1}^{(0)}(a,b)&&=\ -\f 12 (\partial_a+\partial_b){P}_{4K}^{(1)}(a,b)
\nonumber\\
(4K\!-\!2)\ {P}_{4K-3}^{(0)}(a,b)&&=\ -\f 12 \left\{\partial_a\partial_b{P}_{4K-1}^{(0)}(a,b)+ (\partial_a+\partial_b){P}_{4K-2}^{(1)}(a,b)\right\}
\nonumber\\
(4K\!-\!4)\ {P}_{4K-5}^{(0)}(a,b)&&=\ -\f 12 \left\{\partial_a\partial_b{P}_{4K-3}^{(0)}(a,b)+ (\partial_a+\partial_b){P}_{4K-4}^{(1)}(a,b)\right\}
\nonumber\\
&&\cdots. \la{Homog1}
\end{eqnarray}
Hence, degree by degree, all homogeneity components of
${P}_{4K-3}^{(0)}(a,b)$ are determined from those of ${P}^{(1)}(a,b).$

\subsubsection{Second family (forbidden)}

Let us now consider the second family defined by the Ansatz \eqref
{Laplace2}  corresponding to the multipole at $\lam=2.$ The first
equation of \eqref{Eqns} would give
\begin{eqnarray}
\label{Plam2}
Q^{(1)}_{k}(u)=\int_{\circlearrowleft}\frac{d\lambda}{2\pi
i}\ e^{\lambda u}\,
\frac{\Gamma(2K+1)}{\Gamma(2k\!+\!1)[\lambda(\lambda\!-\!2)]^{K-k}}\ \times \f 1{\lam\!-\!2}\ \equiv \frac{\Gamma(2K\!+\!1)}{\Gamma(2k\!+\!1)\Gamma(K\!-\!k\!+\!1)} \ e^{2u} \left[\partial^{K-k}_\lam\left\{\f {e^{\lam u}}{(\lam\!+\!2)^{K-k}}\right\}\right]_{
\lam=0}\ ,
\end{eqnarray}
where the second equality  is  obtained by the change of variable
$\lam\to\lam+2.$ Hence now $Q^{(0,1)}_{k}(u)\equiv e^{2u}\hat
P^{(0,1)}$ are to be taken as products of $e^{2u}$ by a
polynomial.

Now the second equation of \eqref{Eqns} reads \ba
\left\{a\partial_{a}+b\partial_{b}+1
+\frac{1}{2}\partial_{a}\partial_{b}\right\}[e^{2u}\hat
P^{(0)}]=-\f12\{\partial_{a}+\partial_{b}\}[e^{2u}\hat P^{(1)}]\ .
\la{secoQ} \ea
 After differentiating both sides of \eqref{secoQ} and simplifying by the common exponential factor, one gets
\ba \left\{(a+2b)\partial_{a}+(b+2a)\partial_{b}+1
+\frac{1}{2}\partial_{a}\partial_{b}+4(a+b)^2\right\}\hat
P^{(0)}=-\f12\{4(a+b)+\partial_{a}+\partial_{b}\}\hat P^{(1)}\ ,
\la{secohatP} \ea where the additional terms with respect to the
initial second equation \eqref{Eqns} come from the exponential.

Again, as in Eqns. \eqref{Homog1} but with more terms, there
exists {\it a priori} a hierarchy of equations allowing to
determine, term by term, the homogeneous components $\hat
P_d^{(1)}$ from those of $P^{(1)}.$ One obtains
\begin{eqnarray}
4(a+b)^2\ {\hat P}_{4K-1}^{(0)}&&=\ -2 (a+b){\hat P}_{4K}^{(1)}
\nonumber\\
4(a+b)^2\ {\hat P}_{4K-3}^{(0)}&&=\ -\{4K \ +\ 2b\partial_a+2a\partial_b\}{\hat P}_{4K-1}^{(0)}-2(a+b){\hat P}_{4K-2}^{(1)}-\f 12(\partial_a+\partial_b){\hat P}_{4K}^{(1)}
\nonumber\\
4(a+b)^2\ {\hat P}_{4K-5}^{(0)}&&=\ -\{4K\!\!-\!2 \!+\! 2b\partial_a\!+\! 2a\partial_b\}{\hat P}_{4K-3}^{(0)}-
2(a+b){\hat P}_{4K-2}^{(1)}-\frac 12(\partial_a\partial_b)
{\hat P}_{4K-1}^{(0)}-\f 12(\partial_a+\partial_b){\hat P}_{4K}^{(1)}
\nonumber\\
4(a+b)^2\ {\hat P}_{4K-7}^{(0)}&&=\ -\{4K\!\!-\!4 \!+\! 2b\partial_a\!+\! 2a\partial_b\}{\hat P}_{4K-5}^{(0)}-
2(a+b){\hat P}_{4K-4}^{(1)}-\frac 12(\partial_a\partial_b)
{\hat P}_{4K-3}^{(0)}-\f 12(\partial_a+\partial_b){\hat P}_{4K-2}^{(1)}
\nonumber\\
&&\cdots \la{Homog2} ,
\end{eqnarray}
where it clearly appears that at each step the left hand side is
determined by all previously obtained coefficients. However, the
key difference with the set of equations \eqref{Homog1} is that
one has, at each step to factor out the $(a+b)^2$ factors, in
order to ensure the polynomial nature of  $P^{(0)}(a,b)$. Our
conjecture, based on various trials, is that  Eq. \eqref{Homog2},
contrary to the previous case, leads to an obstruction when
reaching the lower homogeneity degrees.

As a simple but significant example, let  us consider  the solution of \eqref{Plam2} with $K=1,$ namely
\ba
\hat P^{(1)}(a,b)\equiv  \sum_{p=0}^{2} \hat P_{2p}^{(1)}(a,b)=(a^2-b^2)^2 + a^2+b^2-1/2\ .
\la{Ex}
\ea
The system \eqref{Homog2} leads to
\begin{eqnarray}
4(a+b)^2\ {\hat P}_{3}^{(0)}(a,b)&&=\ -2 (a+b)^3(a-b)^2
\nonumber\\
4(a+b)^2\ {\hat P}_{1}^{(0)}(a,b)&&=\ - (a+b)(3a^2+3b^2-2ab)
\la{ExHomog2} .
\end{eqnarray}
We see that, while the first equation can be satisfied by
polynomials set, the second cannot. Unless exceptional cases, which
we did not encounter in our tests, there exists an obstruction to
realize the system \eqref{Homog2}. A more general  study of this
mathematical conjecture would be interesting.

\section{Solution and properties}

Let us discuss in detail our results. As we saw, our resulting
basis of regular solutions of the Khalatnikov equation
\eqref{Khalatnikov}, under the form   \eqref {KhalZ}, reads
\begin{equation}
F_K(a,b)=\left[P^{(0)}(a,b|K)\ F^{(0)}(a,b)
 +P^{(1)}(a,b|K)\ F^{(1)}(a,b)\right],
 \label{Znm1}
\end{equation}
where the change $P^{(0,1)}\to P^{(0,1)}(a,b|K)$ identifies the obtained solutions and
\begin{eqnarray}
&&F^{(0)}(a,b)= e^{(a+b)^2}
 \label{Z01}\\
 &&F^{(1)}(a,b)=\int_0^{a+b}e^{t^2}dt\ .
 \label{Z11}
\end{eqnarray}
$P^{(0)}(a,b|K)$, ($resp.\ P^{(1)}(a,b|K)$) span a family of polynomials,
each one  indexed by a different integer $K \in \mathbb{N},$ its
higher degree being $d=4K$ ($d=4K-1$). Note That $P^{(0)}(a,b|0)=1,\ P^{(1)}(a,b|0)=0.$

The following questions are
in order:
\bit \ii {\it How to reconstruct the flow from the
solutions of the Khalatnikov equation?}
\eit
 The guiding line for
constructing the hydrodynamic flow  solutions in the hodographic
method is to obtain the kinematic variables from the solution of
the Khalatnikov potential. The corresponding relation \eqref{zpzm}
expressed using the variables $a,b$ writes \ba z^{\pm}=
\f{1-c^2_s}{8T_0c^2_s}\ \exp{\left\{\f
{2c_s}{1\!-\!c^2_s}\left[c_s(a^2+b^2)\pm
(a^2-b^2)\right]\right\}}\times\left\{ \left(\f\partial{\partial
a^2}+ \f\partial{\partial b^2}\right)\pm c_s
\left(\f\partial{\partial a^2}- \f\partial{\partial
b^2}\right)\right\}
 \chi(a,b)\ .
\la{Reco}
\ea
Now, the physical requirement we are using is the regularity of the
entropy distribution. Its expression is obtained through rewriting the relations (\ref{PhiChi},\ref{Entr}) as the hierarchy of relations starting from the Khalatnikov potential $\chi,$
 \begin{eqnarray}
\Phi&=&\chi(a,b)+\f{g\!-\!1}2 \left\{\partial_{a^2}\partial_{b^2}
\right\} \chi(a,b) \nn \frac{dS}{dy} &=&\frac{s_0(g-1)}{8T_0g}\
e^{-2(a^2+b^2)}\ \{\partial_{a^2}\partial_{b^2}\} \Phi (a,b)\
,\la{RegS}
\end{eqnarray}
where we make use of the relation\begin{equation}
\{\partial_{a^2}+\partial_{b^2}\}\ F(a,b)=\partial_{a^2}\partial_{b^2}\ F
\ ,
 \label{Znm2}
\end{equation}
valid for any solution of  the Khalatnikov equation  $F(a,b)$
 by inserting the definition $Z(a,b)=e^{-(a^2+b^2)}\
F(a,b)$ into \eqref{KhalZ}. Note that the operator
$\{\partial_{a^2}+\partial_{b^2}\}\equiv
\partial_{a^2}\partial_{b^2}$   corresponds exactly to the shift
$K\to K-1$ acting on our basis of solutions \eqref{Znm1}.

From Eqns.(\ref{RegS}) we  see that  the entropy distribution
${dS}/{dy}$ is obtained from up to two successive actions of
$\{\partial_{a^2}\partial_{b^2}\}$ on the Khalatnikov potential
$\chi.$ Hence, in order to obtain a regular ${dS}/{dy}$ it is
required to start with a Khalatnikov  potential $\chi$ at level $K
\ge 2.$ One obtains  the full solution given by
\begin{eqnarray}
\chi&=&\sum_{K_i=2}\lambda_{K_{i}}\ F_{K_{i}}\nn
\Phi&=&\sum_{K_i=2}\lambda_{K_{i}}\ \left(F_{K_{i}}+\f{g\!-\!1}2F_{K_{i-1}}\right)\nonumber\\
\f {dS}{dy} &=& \frac{s_0(g-1)}{8T_0g}\ e^{-2(a^2+b^2)}\ \sum_{K_i=2}\lambda_{K_{i}}\ \left(F_{K_{i-1}}+\f{g\!-\!1}2F_{K_{i-2}}\right) .
\la{Full}
\end{eqnarray}

\begin{figure}[ht]
\begin{center}
\mbox{\epsfig{figure=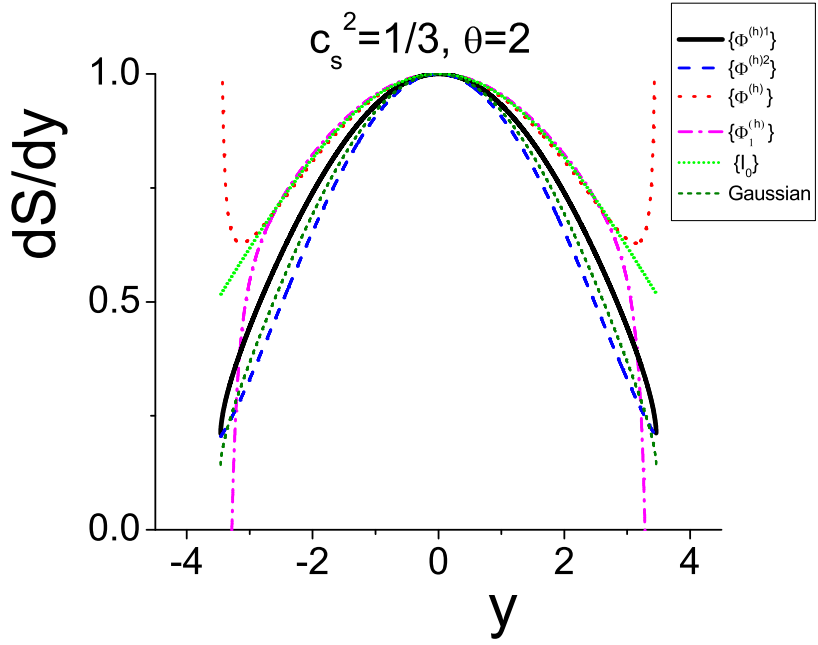,width=12cm,angle=0}} \caption{ {\it
Entropy distribution.} Various  solutions for ${dS}/{dy}$ arising
from the initial harmonic potential  $\Phi^{(h)}$ (see \eqref
{PhiHarm0}) are represented for the speed of sound  $c_s=1/\sqrt
3\ (g=3)$ and $\theta=\log T_0/T=2$. \  {\it Regular solutions:}
1) $\{\Phi^{(h)1}\}$,  solid (black) line : first integral of
$\Phi^{(h)}.$ \ 2) $\{\Phi^{(h)2}\}$, dashed (blue) line : second
integral of $\Phi^{(h)}$. \ {\it Singular solutions:} 1)
$\{\Phi^{(h)}\}$, dotted (red) line : harmonic potential; 2)
$\{\Phi^{(h)}_1\}$, dashed-dotted (magenta) line: first
derivative. \ {\it ``Landau flow'' solutions:}\
 1) $\{Gaussian\}$,  short-dashed (green)
: ``Landau flow'' asymptotics; 2) $\{I_0\}$  thin  (green) line :
Landau-Belenkij exact solution \cite{bi}.
 } \label{g3th2}
\end{center}
\end{figure}

In order to illustrate this discussion, in
Fig.~\ref{g3th2} we present the entropy rapidity-distribution
$dS/dy$ for the first components of the  new basis of solutions.
In particular, we depict $dS/dy$ for the  first two irregular components (starting with the harmonic potential), and for the first two
regular components. Additionally, for completeness we present also
the well-known ``Landau  flow'' solutions, namely the asymptotic Gaussian one \cite{Lan1} and the exact solution of Ref.\cite{bi}.

\bit \ii {\it Is the family identical to the one
obtained in section IIIB by multiple integrations?} \eit
 The
uniqueness of our solutions, up to  overall multiplicative and
additive  constants, for a given maximal degree $4K$ gives a
strong argument that it corresponds exactly to the solutions
obtained by successive integrations over $a^2,b^2.$ In order to
give a simple example, one gets for $K=1$
\begin{eqnarray}
{P}^{(1)}(a,b|1)&&=(a^2-b^2)^2 - a^2-b^2-\f 14 \equiv -\f
14\Pi^{(1)}_{1}
\nonumber\\
{\hat P}^{(0)}(a,b|1)&&=\ -\f 12 (a+b)(a^2-b^2) +\f 14(a+b)\equiv
-\f 14\Pi^{(0)}_{1}\ , \la{Ex1}
\end{eqnarray}
where the polynomials $\Pi^{(0,1)}_{1}$ were obtained in section
IIIB from first integration, see (\ref{int1}). So, the $K$-indexed
basis of solutions is eventually identical to the multiple
integration of the ``harmonic family''. A rigorous mathematical
proof deserves  some more further effort.
 \bit \ii {\it Is the
$K$-indexed family forming a complete basis?} \eit
 This question
appears to be more involved. A well-known example  of solution to
\eqref{KhalZ} (see $e.g.$
\cite{Beuf} for a complete discussion) is given by
\ba Z_0(a,b)= I_0(2ab)=
\partial_{a^2}\partial_{b^2} I_0(2ab)=\partial_{a^2}\{ a/b\times  I_1(2ab)\} \ ,
\label{KhalI}
\ea
where $I_{0,1}$ are the well-known Modified Bessel Functions.
It is clear that those functions cannot be put into the form
\eqref{Znm1} for a combination of solutions with maximal finite
$K_{max}.$ The only possibility would be that $I_{0,1}(a,b)$
belong to the {\it closure} of the basis \eqref{Znm1}, $i.e.$
expressed as a convergent expansion over the basis with $K\to
\infty.$ This example, and more generally the proof of
the completeness of the basis \eqref{Znm1} appears to be a non-trivial mathematical problem which deserves to
be studied on its own. \bit \ii {\it How to introduce the initial
conditions?} \eit The equation \eqref{Znm1} appears to propose a
rich possibility of solutions to the flow equations. However, one
would be interested to modulate these solutions as a function of
the boundary conditions, at least the initial ones. For this sake,
one would have to solve the Green functions \cite{Beuf}, namely
the solutions of the equation
\begin{equation}
c^2_s\,\partial_\theta^2
G(\theta,y)+\left[1-c_s^2\right]\partial_\theta
G(\theta,y)-\partial^2_y G(\theta,y)=\dl(\theta)\dl(y)\ . \ea We
hope our method could be useful to solve in the near future this equation.

\section{Conclusion and Outlook}

Let us summarize our  results:

 i) We have derived a basis of  solutions for the $(1\!+\!1)$\,-dimensional flows of a perfect fluid with arbitrary constant speed of sound. It
spans an infinite-dimensional linear vectorial space of solutions.

 ii) The basis elements can
be indexed by a positive or negative integer number $K\in {\cal Z},$ where the negative indices
$K <0$ correspond to {\it singular} solutions while
$K\in {\cal N}$ correspond to {\it regular} ones.

 iii) The singular basis can  be obtained by successive differentiation
of the ``harmonic flow'' solution, while the regular one arises
from successive integration.

iv) The general {\it regular} solution is characterized by a Khalatnikov potential which is an arbitrary combination of components of a linear basis
\begin{equation}
\{\chi(a,b|K)\} = \left\{P^{(0)}(a,b|K)\ e^{(a+b)^2}
 +P^{(1)}(a,b|K)\ \int_0^{a+b}e^{t^2}dt\ \ \left\vert\ K \in [2,\infty] \right.\right\}\ ,
 \label{Sol1}
\end{equation}
where $P^{(1,0)}(a,b|K)$ are polynomials of maximal homogeneity
degree  $d=(4K,4K\!-\!1)$ defined by recurrence relations, see
formulas (\ref{Plam1},\ref{Homog1}). They are uniquely defined (up
to an overall constant) by the value of $K.$

v) The variables describing the flow
 \ba
a\equiv\frac{\sqrt{|\theta|+\frac{y}{\sqrt{g}}(g\!-\!1)}}{2}\ ,\ \
\ b\equiv\frac{\sqrt{|\theta| -\frac{y}{\sqrt{g}}(g\!-\!1)}}{2}
 \label{Sol2}
\ea
are functions of the dynamical variables $\theta=\log T_0/T$, $y=\f 12 \log u^+/u_-$ depending on the flow temperature $T$ and longitudinal velocity $u_\pm=u_0\pm u_1$. The kinematics of the flow are recovered from the Khalatnikov potential through the inverse hodograph transformation $(\theta,y)\to z_\pm =z_0\pm z_1$.

As an outlook, let us quote a few interesting problems following our present study:

- Can we prove (or disprove) the completion of our vectorial space
of regular solutions or, equivalently,  can we prove (or disprove)
that any regular Khalatnikov potential solution arises from a
convergent expansion
$\sum_{K=2}^{\infty}\lambda_{K_i}\chi(a,b|K_i)$?

- Can we solve the solution with given boundary conditions or,
equivalently, solve the Green function associated with the
Khalatnikov equation?

- Can we strictly satisfy the energy conservation inside the
forward light-cone? This arises from  the remark \cite{Janipri}
that the harmonic flow does not verify this constraint.

We hope that the presented solution for the rather old problem of
$(1\!+\!1)$\,-dimensional  flows of a perfect fluid can serve for
the modern and stimulating hydrodynamic investigations both on the
phenomenological (through heavy-ion experiments) and theoretical
(through Gauge/Gravity duality) points of view. In any case, it
was quite pleasant to deal with this problem.
\\

\paragraph*{{\bf{Acknowledgements:}}}
Emmanuel N. Saridakis wishes to thank Institut de Physique
Th\'eorique, CEA, for the hospitality during the preparation of
the present work.

\end{document}